\documentclass[twocolumn]{jpsj3} %% two-column layout
\usepackage{color}
\usepackage{multirow}
\usepackage{graphicx}  % Include figure files
\usepackage{dcolumn}   % Align table columns on decimal point
\usepackage{bm}        % bold math
\usepackage{color}

\newcommand{\G}{\Gamma}
\newcommand{\dg}{\dagger}
\newcommand{\up}{\uparrow}
\newcommand{\dw}{\downarrow}

\newcommand{\be}{\begin{eqnarray}}
\newcommand{\ee}{\end{eqnarray}}
\newcommand{\s}{\sigma}

\def\simge{\lower0.7ex\hbox{$\ \overset{>}{\sim}\ $}}
\def\simle{\lower0.7ex\hbox{$\ \overset{<}{\sim}\ $}}

\title{Local Nodal Cooper Pairs in Multiorbital Systems}

\author{Kazumasa Hattori$^{1}$\thanks{E-mail:
hattori@tmu.ac.jp} 
Takuya Nomoto$^2$, Takashi Hotta$^1$, and 
Hiroaki Ikeda$^3$}
\inst{$^1$Department of Physics, Tokyo Metropolitan University, 
Minami-osawa, Hachioji, Tokyo 192-0397, Japan\\
$^2$RIKEN Center for Emergent Matter Science (CEMS), Hirosawa, Wako, Saitama 351-0198, Japan\\
$^3$Department of Physics, Ritsumeikan University, Kusatsu, 525-8577, Japan%
}

\abst{We show the occurrence of a new class of superconductivity in multiorbital systems, focusing on non-Kramers f$^2$ states. The Cooper pairs in this class of superconductivity are mainly local pairs with the same symmetry as the local f$^2$ ground states. When the local ground state is an anisotropic representation, the superconducting gap has nodes on the Fermi surface. This nodal superconductivity is mediated by the strong on-site interorbital attractions arising from the negative-$U$ physics, generalized in multiorbital systems. We show that this is realized in a simple two-orbital model with antiferro Hund's coupling and enhanced inter-orbital interactions derived via a systematic local down folding. Finally, we briefly discuss superconductivity in Pr-1-2-20 compounds, UBe$_{13}$, and PrOs$_4$Sb$_{12}$, in view of the present mechanism.
}
\usepackage{amsmath}	% required for `\align' (yatex added)
\begin{document}
\maketitle

%%%%%%%%%%%%%%%%%%%%% Introduction %%%%%%%%%%%%%%%%%%%%%%%%%%%%%%%%%%%%%%%%%%%%
Unconventional superconductivity (SC) shows various interesting phenomena and has attracted great attention 
in the field of condensed matter physics. The existence of nodes in their superconducting gap functions is required for 
these phenomena to occur in unconventional superconductors, such as cuprates,\cite{highTc} ruthenates,\cite{SrRu} iron-based 
pnictides,\cite{Fe} and heavy-fermion superconductors.\cite{heavy}

Apart from cuprates and some others, the gap functions in many unconventional superconductors are not fully understood and continue to be under debate, despite the intensive experimental and theoretical studies conducted on them since their discovery. Thus, explaining their mechanism is a challenging problem in condensed matter theory.

A promising mechanism for unconventional SC in single band (orbital) systems, fluctuation-mediated SC,  
has been established by the 80th,\cite{Miyake} and is analogous to the theory of $^3$He superfluids.\cite{He}
In particular, intersite fluctuations in the presence of strong local repulsions, such as ferro- or antiferro-magnetic fluctuations, 
lead to nodal SC. Many unconventional superconductors have been observed in close vicinity to 
ordered phases.

In recent years, much attention has been paid to multiorbital superconductors, such as iron-based pnictides.\cite{Fe} 
Some heavy-fermion superconductors have attracted renewed interest, since for example, the discovery of full-gap behavior in the low-temperature specific heat of CeCu$_2$Si$_2$.\cite{Kittaka} 
Thus, it is important to clarify the impact of the orbital degrees of freedom on SC.
Recently, to clarify the multiorbital character of such SC, we classified multipole SC\cite{Nomoto} and discussed that nodal SC can occur through the formation of local Cooper pairs in multiorbital systems. The pairs are local but have orbital degrees of freedom, which form the nodal gap structure.

In this letter, we show that such nodal and local SC can emerge in multiorbital systems and that it is related to two-electron ground state configurations, when the electron filling is nealy  two per site. We start first by demonstrating that low-energy effective interactions in such multiorbital models with the spin-orbit interaction (SOI) are 
completely different from the conventional Hubbard-type parametrization, and include e.g. antiferro Hund's coupling\cite{Yotsuhashi} and enhanced interorbital interactions.
%%%%%%%%%%%%%%%%%%%%%%%%%%% fig 1 %%%%%%%%%%%%%%%%%%%%%%%
\begin{figure}[b]
\vspace{-5mm}
\begin{center}
\includegraphics[width=0.35\textwidth, angle=270]{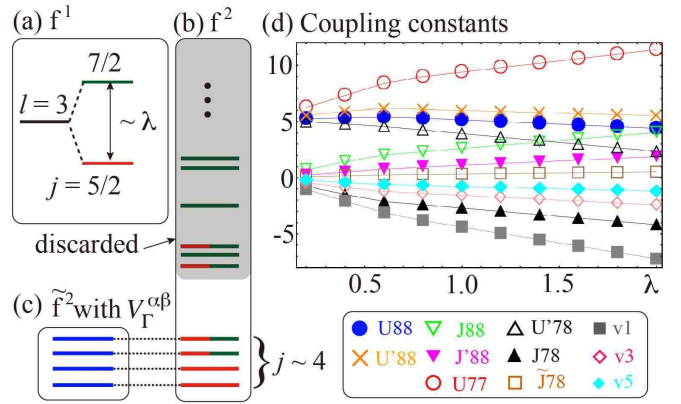}
\end{center}
%\vspace{-0.5cm}
\caption{
(Color online) Schematic energy spectra for the (a) f$^1$, (b) f$^2$ states, and (c) the low-energy effective f$^2$ states, denoted as $\tilde{\rm f}^2$. (d) Coupling constants for the effective $\tilde{\rm f}$ system as a function of $\lambda$ when $x=0$, $W=0.004$, and $(F_0,F_2,F_4,F_6)=(9,6,5,2)$. The unit of energy is eV.
}
\label{fig-3}
\end{figure}
%%%%%%%%%%%%%%%%%%%%%%%%%%%%%%%%%%%%%%%%%%%%%%%%%%%%%%%%%%

Renormalized interactions after 
integrating the high-energy sectors contain important information for understanding the low-energy properties of complex systems.
In this work, we show a typical example in $f$-electron (where the orbital angular momentum $l=3$) systems under the $O_h$ group, 
demonstrating the effective local interactions in $j=l-1/2=5/2$ multiplets, 
by integrating the $j=l+1/2=7/2$ states that lie in the higher energy of 
the SOI order, 
 $\lambda$ 
%$\sim \frac{7}{2}\lambda$. 
(See Fig. \ref{fig-3}(a)). This is a kind of down folding,\cite{downfolding} which we will call local down folding.

A Hubbard-type model for the $j$ $ = $ $5/2$ orbital has already been analyzed, and it shows that realization of the $\G_3$ nonmagnetic Kramers doublets as ground states for the local two-electron sector (f$^2$) is not possible, when considering the conventional Slater integrals $F_{0,2,4,6}$, among the $j=5/2$ orbitals.\cite{Hotta}  
Thus, when analyzing $\G_3$, as realized in many Pr- and U-based compounds\cite{PrU} that show quadrupole Kondo effects\cite{Cox} and orbital orders, \cite{PrU}
a simple $j=5/2$ Hubbard-type model is insufficient.
To overcome this limitation, $1/\lambda$ perturbative analysis has been conducted thus far.\cite{Hotta2} 
In this letter, we show a more rigorous formulation suitable for the realistic parameters (i.e., $\lambda\ll$ Hund's coupling),  to analyze f$^2$-based materials with $\G_3$ ground states. 
To this end, we use a method developed in the contractor renormalization group,\cite{core} 
and obtain the local effective interactions among
 the low-energy fermions, which are, roughly speaking, $j=5/2$
 states in the case where the filling is nearly two per site.

The procedure for calculating the renormalized interactions is as follows: 
(i) The exact diagonalizations are carried out for the local Hamiltonian including both $j=5/2$ and $7/2$, with the Coulomb interactions $F_{0,2,4,6}$, and the SOI 
 under the crystalline-electric-field (CEF) potential expressed by the parameters $x$ and $W$.\cite{TraditionalNotation} Subsequently, the eigenenergy $E^{(n)}_{\Gamma,s}$ for the f$^n$ configurations and the corresponding wavefunctions $|{\rm f}^n\Gamma, s\rangle$, where $\G$ represents the irreducible representation (irrep) for the $O_h$ group and $s$ runs $1,2,\cdots$ with increasing energy are obtained. In this work, only the information on $n\le 2$ is needed. It should be noted that the low-energy f$^2$ states are $j \sim 4$, as predicted by Hund's rule.
(ii) The target low-energy f$^1$ states are set, which belong to $\Gamma_7$ or $\Gamma_8$, while the high-energy f$^1$ states are ignored.
 For realistic parameters, they are almost $j=5/2$ states and their creation operators are denoted as $\tilde{f}^{\dag}_{\Gamma_{7,8}}$. 
(iii) The f$^2$ states are constructed as $|\tilde{\rm f}^2\Gamma^{(\alpha)}\rangle=(\tilde{f}^{\dag}\tilde{f}^{\dag})_{\Gamma^{(\alpha)}}|0\rangle$, 
where the states for $\alpha=1$ consist of two $\G_8$ orbitals, while that for $\alpha=2$ includes at least one $\G_7$ orbital:\cite{wavefunc} $\Gamma^{(\alpha)}=\Gamma_1^{(1,2)}$, $\Gamma_3^{(1,2)}$, $\Gamma_4^{(1)}$, or $\Gamma_5^{(1,2)}$. 
(iv) The overlap $r^{\alpha s}_{\Gamma}=\langle \tilde{\rm f}^2\Gamma^{(\alpha)}|
{\rm f}^2\Gamma,s\rangle$ is calculated. 
(v) The effective Hamiltonian for $\tilde{f}$ is set as:
\be
H_{\rm eff}=\sum_{m=7}^8E^{(1)}_{\G_m,1}\tilde{f}^\dag_{\Gamma_m}\tilde{f}_{\G_m}+\sum_{\Gamma,\alpha\beta} V^{\alpha\beta}_{\Gamma}|\tilde{\rm f}^2\Gamma^{(\alpha)}\rangle\langle \tilde{\rm f}^2\Gamma^{(\beta)}|,  \label{Heff}
\ee
where 
$V_{\G}^{\alpha\beta}$ are determined so as to (a) maximize the overlap between 
 $|{\rm f}^2\Gamma, 1\rangle$ and the ground state of $H_{\rm eff}$ for each $\Gamma$: $|\tilde{\rm f}^2\Gamma\rangle_{gs}$ and to (b) reproduce the
 $E^{(2)}_{\Gamma, s}$ for the first (two) $s$ with nonzero $r^{\alpha s}_{\Gamma}$'s for $\Gamma=\Gamma_4 (\Gamma_1, \Gamma_3, \Gamma_5)$.
The condition for maximizing the overlap leads to 
$|\tilde{\rm f}^2\Gamma\rangle_{gs}\propto \sum_{\alpha}r^{\alpha 1}_{\Gamma}|\tilde{\rm f}^2\Gamma^{(\alpha)}\rangle$. The $H_{\rm eff}$ constructed reproduces exactly the local spectra up to two-electron filling.
To reproduce the spectra of f$^3$, f$^4$, and $\cdots$, three- and four-body interactions and so on need to be included. 
When the interest is only in the f$^{0,1,2}$ states, they can be ignored.

Symmetry reduces the number of independent 
parameters 
$V^{\alpha\beta}_{\Gamma}$ to ten.\cite{opRenormalization} 
The three parameters are those in the $\G_8$ sector: 
$U_{88}, U'_{88}$, $J_{88}$, and $J'_{88}$ with the cubic constraint $U_{88}=U'_{88}-3J_{88}/4+J'_{88}$ in $V_{88}$, where 
\begin{eqnarray}
V_{88}&=&U_{88}(n_{a\up}n_{a\dw}+n_{b\up}n_{b\dw})
+U^{\prime}_{88}n_{a}n_{b}
+J_{88}{\bf S}_{a}\cdot{\bf
S}_{b}\nonumber\\
&&+J^\prime_{88}\big(a_{\up}^{\dg}a_{\dw}^{\dg}b_{\dw}b_{\up}+{\rm H.c.}\big).  \label{Hlocal}
\end{eqnarray}
Here, the annihilation operator for the $\Gamma_8$ orbital is denoted 
as $\{a_{\up},a_{\dw},b_{\up},b_{\dw}\}(\equiv \psi)$, where $a$ and $b$ represent 
the two kinds of orbital degrees of freedom and $\sigma=\up, \ \dw$ represents 
the Kramers index. $n_{a\s}$$=$$a_\s^\dg a_\s$, $n_{a}=\sum_{\s} n_{a\s}$ and ${\bf S}_a=\frac{1}{2}\sum_{\s\s'}a^\dg_{\s} 
\vec{\s}_{\s\s'}a_{\s'}$ with $\vec{\s}$$=$$(\s^x, \s^y, \s^z)$ being the Pauli matrices and similar expressions for the $b$ orbital are used. 
By denoting the $\G_7$ creation operator as $c_{\sigma}^{\dag}$, with $n_7=\sum_{\s}c_\s^\dag c_\s$  and 
${\bf S}_7=\frac{1}{2}\sum_{\s\s'}c^\dg_{\s} 
\vec{\s}_{\s\s'}c_{\s'}$,
the other parts are obtained as:  
\begin{eqnarray}
V_{78}\!\!\!\!\!\!\!\!\!\!\!\!&&=U_{77}c_\up^\dag c_\up c_\dw^\dag c_\dw+U'_{78}n_{7}n_8+J_{78}{\bf S}_7\cdot {\bf S}_8\nonumber\\
\!\!\!\!\!\!\!\!\!\!\!\!\!\!\!\!\!\!\!\!\!&&+\tilde{J}_{78}
(\tau^z,\tau^x) \cdot 
\begin{pmatrix}
2S^z_8S_7^z-S_8^xS_7^x-S_8^yS_7^y\\
\sqrt{3} (S_8^xS_7^x-S_8^yS_7^y)\\
\end{pmatrix}\nonumber\\
\!\!\!\!\!\!\!\!\!\!\!\!\!\!\!\!\!\!\!\!\!
&&+\big[v_1(\psi^\dag \psi^\dag)_{\G_1}c_{\up}c_{\dw}\!+\!\!\sum_{m=3,5}\!\!v_m(\psi^\dag \psi^\dag)_{\G_m}(c\psi)_{\G_m}
\!\!+\!{\rm H.c.}\big],\ \ \ \ \ \label{Heff2}
\end{eqnarray}
where  $n_8=n_a+n_b$,  ${\bf S}_8={\bf S}_a+{\bf S}_b$, and $\vec{\tau}=(\tau^z,\tau^x)$ are the Pauli matrices for the orbital indices $a(b)\to \up(\dw)$.

Figure \ref{fig-3}(d) shows various $V^{\alpha\beta}_{\G}$ as a function
of $\lambda$ for $x=0$, $W=0.004$ eV, and $(F_0,F_2,F_4,F_6)=(9,6,5,2)$ eV. For real
materials, the value of $\lambda$ is expected to be 
$\lambda\simle 0.5$ eV. The results are summarized as follows:  

\begin{itemize}
\item $J_{88},J^\prime_{88}>0$ and they increase as $\lambda$ increases. 
%The largest coupling in the magnitude is $U_{77}$ and, 
\item $U^\prime_{88}>U_{88}$, i.e., the interorbital interaction is larger than 
the intraorbital interaction for the $\G_8$ orbital. 
\item The complex pair hopping terms (in particular for $v_1$) are quite large in magnitude. 
\end{itemize}

{\flushleft It should also be noted that the exchange interaction $J_{78}$ between $\G_7$ and $\G_8$ is ferromagnetic.
In addition to the purely electronic origin for such parametrization of the interactions, 
electron-phonon couplings are known to enhance e.g., antiferro Hund's coupling.\cite{Nomoto, Hoshino2}  }

%\subsection{Pairing interactions and 2-body CEF states in cubic $\Gamma_8$ orbital models.}
Now, we discuss that the above results affect the ordering and SC. To capture the essential points, we simplify the full $j=5/2$ model and introduce an effective model with a four-fold degenerate $\Gamma_8$ orbital. Although the model is an effective one, and thus, should be regarded as an $\tilde{\rm f}$ system with regard to eq. (\ref{Heff}); notations such as f$^n$ will be used for simplicity. 
As discussed, representing the local interactions for the $\Gamma_8$ orbital as
$V_{88}$ from eq. (\ref{Hlocal}), 
the f$^2$ configurations are diagonalized as $\G_1$(1-fold), $\G_3$(2-fold), and $\G_5$(3-fold). 
Apart from the $\G_8$ level, their atomic-limit energy $\epsilon^{(2)}_{m} (m=1,3,5)$ is given as 
$\epsilon^{(2)}_{1}=U_{88}+J^\prime_{88}$, $\epsilon^{(2)}_3=U_{88}-J^\prime_{88}$, and $\epsilon^{(2)}_5=U_{88}+J_{88}-J^\prime_{88}$. 
Thus, the f$^2$ ground state is determined by $J_{88}$ and $J^\prime_{88}$. 

We are interested in the situation where the filling is $n\equiv\langle n_8 \rangle\sim 2$ per site and in the SC arising there. In a naive mean-field approximation, the $U_{88}$ term in eq. (\ref{Hlocal}) can be decoupled into 
 a density-density form, while the other terms can be decoupled into Cooper channels. 
Thus, the interaction (\ref{Hlocal}) can be rewritten as: 
$
V_{\rm loc}=U_{88}n(n-1)/2+\sum_{m=1,3,5}(\epsilon^{(2)}_{m}-U_{88})(\psi^{\dg}\psi^{\dg})_{\Gamma_{m}}(\psi\psi)_{\Gamma_{m}}
$, 
where $(\psi^\dg\psi^\dg)_{\G_m}$ indicates the two-electron operator for $\G_{m}$ 
irrep.\cite{wavefunc} 
A mean-field analysis readily leads to SC with 
local Cooper pairs corresponding to the {\it atomic CEF ground state} in the f$^2$ sector $m=m_g$, as the interaction is always attractive ($\epsilon_{m_g}^{(2)}-U_{88}<0$). 
For this analysis, it is crucial to perform the decoupling of the $U_{88}$ term not in the Cooper channel but in the density-density channel. 
Physically, if the charge fluctuations are suppressed, 
the above analysis is expected to be valid, and the common energy contribution $U_{88}$ to $\epsilon^{(2)}_m$ is irrelevant. 
In fact, in recent dynamical mean-field theory studies on multiorbital  
Hubbard models, realization of local $s$-wave SC for $U^\prime > U > 0$\cite{Koga} and local-triplet SC in a three-orbital model\cite{Hoshino} was shown.
 These results support the validity of the above analysis.
 In the following, we show that such SC can be nodal, if realistic hopping parameters, reflecting the orbital characters, are taken into account.

As an example, a model on a simple cubic lattice is considered.
The non-interacting Hamiltonian is:
\begin{equation}
H_0=\sum_{\bf k}\psi^{\dag}_{{\bf k}\alpha}[\epsilon({\bf k}) +\vec{d}({\bf k})\cdot \vec{\tau} + 
 \vec{\eta}({\bf k})\cdot \vec{\sigma}\tau^y-\mu]_{\alpha\beta}\psi_{{\bf k}\beta}, 
\end{equation}
 where $\mu$ is the chemical potential, $\alpha$ and $\beta$ are run for the indices $\{1,2,3,4\}\equiv\{a_{\up}, a_{\dw}, b_{\up},b_{\dw}\}$, and ${\bf k}$ is the wavenumber. $\vec{\sigma}$ acts on the Kramers indices $\up$ or $\dw$, while $\tau^y$ is the $y$ component of the Pauli matrices for the orbital indices. The Einstein contraction for the repeated indices is used, and will be used hereafter. $\epsilon({\bf k})$, $\vec{d}(\bf k)$, and $\vec{\eta}(\bf k)$ are
real and they transform as $A_{1g}$, $E_{g}$, and $T_{2g}$ in the $O_h$ group,  
respectively,\cite{dvec} where 
the hopping integrals up to the third neighbors are taken into account. In the following analyses, the unit of energy is set to the orbital 
diagonal nearest-neighbor hopping $t=1$ and the unit of length is set to the lattice constant. The other parameters are set to:
$(t',t'',d,d',\eta,\eta')=(0.4,0.1,0.3,0.1,0.1,0.05)$.
The one-particle energy $E_{\lambda=1,2}({\bf k})$  is:
\begin{equation}
E_{\lambda}({\bf k}) =\epsilon({\bf k})+(-1)^\lambda\sqrt{|\vec{d}({\bf
 k})|^2+|\vec{\eta}({\bf k})|^2}-\mu, %\equiv \epsilon({\bf k})+(-1)^\lambda D({\bf k})
\end{equation}
with Kramers degeneracy. The band-based operators are defined as $(\tilde{c}_{{\bf k}1\up},\tilde{c}_{{\bf k}1\dw},\tilde{c}_{{\bf k}2\up},\tilde{c}_{{\bf k}2\dw})^{\rm T}$ $=$ $[U({\bf k})]^\dag (a_{{\bf k}\up},a_{{\bf k}\dw}, b_{{\bf k}\up},b_{{\bf k}\dw})^{\rm T}$, where $U({\bf k})$$=$$U(-{\bf k})$ is unitary and the superscript T indicates the transpose.

Table \ref{tbl-1} summarizes the band-based (intraband) pair amplitudes 
$\tilde{\Phi}_{1,2}({\bf k})=\langle \tilde{c}_{{\bf k}1,2 \uparrow}\tilde{c}_{-{\bf k}1,2\downarrow}\rangle$, induced 
by a local pair amplitude $\Phi_{\G}=\langle (\psi_{\bf i}\psi_{\bf i})_{\G}\rangle$ with $\bf i$ being the site index. For simplicity, interband pairs are not considered here. 
It should be noted that $\tilde{\Phi}_{1,2}({\bf k})=\tilde{\Phi}_{1,2}(-{\bf k})$ due to the local nature of the pair, and 
the symmetry of $\tilde{\Phi}_{1,2}({\bf k})$ is the same as the local order parameter $\Phi_{\G}$.   
Thus, nodal SC is realized when the 
f$^{2}$ ground state is $\G_3$ or $\G_5$. 
The pair is local,\cite{note} and this contrasts with the conventional non-$s$-wave intersite pairs.\cite{Miyake} 
The local nature of the SC suggests that it is robust against detailed changes in the band structure. 
In a recent paper,\cite{Bishop} Bishop et al. discussed such types of SC with nodes, while they introduced attraction in a specific Cooper channel from the beginning. 
Such nodal SC has also been shown for $j=3/2$ fermions in half-Heusler compounds.\cite{Brydon}

\begin{table}[t]
\vspace{-0.2cm}
\caption{Relation between the local and band-based pair
 amplitudes. Trivial constant factors are omitted, 
 $\hat{d}_{z,x}=d_{z,x}({\bf k})/D({\bf k})$ and
 $\hat{\eta}_{x,y,z}=\eta_{x,y,z}({\bf k})/D({\bf k})$, where 
 $D^2({\bf k})\equiv |\vec{d}({\bf
 k})|^2+|\vec{\eta}({\bf k})|^2$ and is invariant under the $O_h$ symmetry.
The abbreviations $c_{x,y,z}\equiv \cos k_{x,y,z}$ and $s_{x,y,z}\equiv
 \sin k_{x,y,z}$ are used. In the third column, the functional form for
 $\Phi_{\G_m}$ is shown (the common factor $1/D({\bf k})$ for $m\ge 3$ is omitted for
 simplicity).} 
\label{tbl-1}
\begin{tabular}{>{\centering\arraybackslash}p{10mm}>{\centering\arraybackslash}p{18mm}>{\centering\arraybackslash}p{47mm}} 
\hline
$\Phi_{\G}$ & $\tilde{\Phi}_\lambda({\bf k})$ & functional form  \\
  \hline \hline
$\Phi_{\G_1}$ & 1 & 1\\
$\Phi_{\G_3,u}$ & $\hat{d}_z({\bf k})$ & $d(2c_z-c_x-c_y)+d'(2c_xc_y-c_yc_z-c_zc_x)$\\
$\Phi_{\G_3,v}$ & $\hat{d}_x({\bf k})$ & $\sqrt{3}d(c_x-c_y)$$+\sqrt{3}d'(c_yc_z-c_zc_x)$\\
$\Phi_{\G_5,xy}$ &  $\hat{\eta}_z({\bf k})$ & $(\eta+\eta'c_z)s_xs_y$\\
$\Phi_{\G_5,yz}$ & $\hat{\eta}_x({\bf k})$ & $(\eta+\eta'c_x)s_ys_z$\\
$\Phi_{\G_5,zx}$ & $\hat{\eta}_y({\bf k})$ & $(\eta+\eta'c_y)s_zs_x$\\
\hline\hline
\end{tabular}
\vspace{-0.5cm}
\end{table}

To examine whether such SC occurs,   
the multiorbital random-phase approximation (RPA)\cite{Takimoto,Takimoto2} is employed and used to calculate the transition temperature $T_{\rm sc}$ of SC and $T_{\rm c}$ for possible multipole orders. 
Generalized static susceptibilities
$\chi_{\gamma\alpha\delta\beta}({\bf q})\equiv N^{-1}
\sum_{\bf kp}\int_0^{1/T}\!\! d\tau\langle T_{\tau}\psi^\dag_{{\bf k}\gamma}(\tau)\psi_{{\bf
k+q}\delta}(\tau)\psi_{{\bf p}\beta}^{\dag}(0)\psi_{{\bf
p-q}\alpha}(0)\rangle$, 
 where $N$, $T$, and $T_{\tau}$ represent the
total number of sites, the temperature, and the time-ordered product,
respectively, are given in the RPA as:
\begin{eqnarray}
 \chi^{\rm RPA}_{\gamma\alpha\delta\beta}({\bf
  q})=\chi^0_{\gamma\alpha\delta\beta}({\bf
  q})-\chi^0_{\gamma\alpha\alpha'\gamma'}({\bf q})
\G_0^{\alpha'\beta'\delta'\gamma'}
\chi^{\rm RPA}_{\beta'\delta'\delta\beta}(\bf q). \label{chiRPA}
\end{eqnarray}
Here,  
$\chi^0_{\gamma\alpha\delta\beta}({\bf q})$$\equiv$$-TN^{-1}\sum_{\omega_\ell\bf k}
G^0_{\beta\gamma}(\omega_\ell,{\bf k}) G^0_{\alpha\delta}(\omega_\ell,{\bf
k+q})$ 
with the Matsubara frequency $\omega_\ell$ and 
$G_{\alpha\beta}^0(\omega_\ell,{\bf k})=\sum_{\lambda\sigma}$$U_{\alpha,\lambda\sigma}({\bf k})U^*_{\beta,\lambda\sigma}({\bf k})$$/$$
[i\omega_\ell$$-$$E_{\lambda}({\bf k})].
$
For convenience, the antisymmetrized interactions $\Gamma_0^{\alpha\beta\delta\gamma}$ are introduced
and eq. (\ref{Hlocal}) is rewritten as:
$V_{\rm loc}$$=$$\frac{1}{4}\G_0^{\alpha\beta\delta\gamma}
\psi_{\alpha}^\dag\psi_{\beta}^\dag\psi_{\delta}\psi_{\gamma}$ with
$\G_0^{\alpha\beta\delta\gamma}$$=$$V_0^{\alpha\beta\delta\gamma}$$-$$V_0^{\alpha\beta\gamma\delta}$$-$$V_0^{\beta\alpha\delta\gamma}$$+$$V_0^{\beta\alpha\gamma\delta}$, 
where $V_0^{\alpha\beta\delta\gamma}$ is given by eq. (\ref{Hlocal}).\cite{nonvanishV} 
The effective RPA interactions $V^{\alpha\beta\delta\gamma}_{\rm eff}$ between the Cooper pairs can be expressed as:
\begin{eqnarray}
V^{\alpha\beta\delta\gamma}_{\rm
 eff}({\bf q})=\frac{1}{2}\Gamma_0^{\alpha\beta\delta\gamma}-\Gamma_0^{\alpha\beta'\delta'\gamma}\chi^{\rm
 RPA}_{\beta'\delta'\alpha'\gamma'}({\bf q})\Gamma_0^{\alpha'\beta\delta\gamma'},\label{Veff}
\end{eqnarray}
which are used for calculating $T_{\rm sc}$ in the BCS approximation. 
 All the calculations shown below are performed for $N=64^3$ and 1024 $\tau$ bins, and $J_{88},J^\prime_{88}>0$, 
which corresponds to the $\G_3$ ground states in the f$^2$ sector.

%%%%%%%%%%%%%%%%%%%%%%%%%%% fig 1 %%%%%%%%%%%%%%%%%%%%%%%
\begin{figure}[t]
\begin{center}
\includegraphics[width=0.4\textwidth]{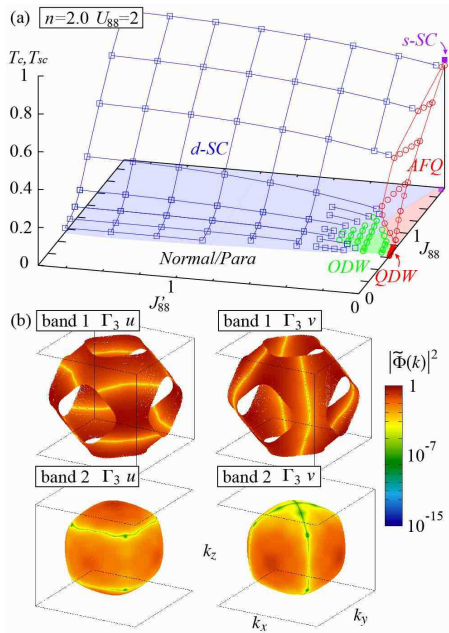}
\end{center}
\vspace{-0.5cm}
\caption{
(Color online) (a) Transition temperatures $T_{\rm sc}$ and $T_{\rm c}$ in the $J_{88}$-$J_{88}^\prime$ plane. (b) Fermi surfaces of band 1 and 2 for $(J_{88},J'_{88})=(0.75,1.0)$ 
and $T\simeq T_{\rm sc}=0.06$. The color map represents $|\tilde{\Phi}_{1,2}({\bf k})|^2$ on the FS. 
}
\label{fig-1}
\vspace{-1cm}

\end{figure}
%%%%%%%%%%%%%%%%%%%%%%%%%%%%%%%%%%%%%%%%%%%%%%%%%%%%%%%%%%
Figure \ref{fig-1}(a) shows the $J_{88}$ and $J^\prime_{88}$ dependence of $T_{\rm sc}$ and $T_{\rm c}$ for $U_{88}=2.0$ and $n=2.0$. It should be noted that 
$T_{\rm sc}$ is not calculated in the ordered phases below $T_{\rm c}$. For a wide range of parameter space,
SC with $E_g$ symmetry ($d$-SC) occurs. This is nearly local SC induced by the renormalized interactions. 
Thus, our naive analysis is qualitatively valid within 
the RPA level. The Fermi surfaces (FSs) for $(J_{88},J'_{88})=(0.75,1.0)$ and $T \sim T_{\rm sc}$ are shown in Fig. {\ref{fig-1}} (b) with the $d$-wave amplitudes $|\tilde{\Phi}({\bf k})|^2$. Line nodes exist on both FSs, whose functional forms are qualitatively consistent with the list in Table \ref{tbl-1}. 
 Below $T_{\rm sc}$, when the chiral combination of the two components is realized 
in favor of opening the gaps on the FSs, point nodes exist around the [111] directions. 
For larger 
$J_{88}$ and $J'_{88}$, $T_{\rm sc}$ is very high: $\sim O(1)$, which means that the attraction is too large to be handled in the RPA. 
Although it is beyond the scope of this study, 
tightly bound pairs dominate and Bose-Einstein condensation (BEC)
of the local pairs would occur.\cite{NSR} 
In such a situation, 
the $\Gamma_3$ bound-state contribution would have to be accounted for, which would suppress $T_{\rm sc}$.
It should be noted that a transition (not a crossover) between SC and BEC 
is expected to occur for $d$-wave pairing.\cite{d-BEC}

  For smaller $J'_{88}$, three ordered states appear: an antiferro $E_g$-quadrupole order (AFQ) with the ordered wavenumber ${\bf q}=(\pi,\pi,\pi)$, and an $E_g$-quadrupole density wave (QDW) and $A_{2g}$-octupole density wave (ODW), both with ${\bf q}\simeq(\pi,\pi,\pi\pm0.4\pi)$ and the equivalent ${\bf q}$'s.

%These two orders (in some sense accidentally) correspond to the two degrees of freedom within the $\G_3$ ground states for the local two-electron sector.  
For $J'_{88}\ll J_{88}$ and $J_{88}\simeq U_{88}$, another type of SC with $A_{1g}$ irrep ($s$-SC) emerges. This is related to the SC realized for $J'_{88}<0$, where the ground state for the f$^2$ sector is $\G_1$. The $\G_1$ ground state, according to Table \ref{tbl-1}, leads to $A_{1g}$ Cooper pairs. With regard to the $\G_5$ f$^2$ ground state with $J_{88}<0$, SC with $T_{2g}$ irrep is expected, while this is not realized for $U_{88}=2.0$, since an antiferromagnetic order takes place  first. However, for larger $|J_{88}|$, the leading SC instability is $T_{2g}$ type,\cite{Hoshino} and thus, it is concluded that the naive approximation used holds true for the RPA results, concerning the dominant SC instability.

%Although the $T_{sc}$ is very high, and thus, the present scheme is 
%insufficient, the mechanism for this SC is characteristic in multiorbital systems with SO coupling; the attraction between the Cooper pairs are not due to isotropic %interactions\cite{Nomoto} but anisotropic ones allowed in the simple cubic lattice. 
%For other part, one notices that SC with $A_{2u}$ symmetry appears for $J^\prime_{88}\ll J_{88}$ 
%and this SC is {\it conventional} fluctuation-mediated SC as 
%anticipated close to some ordered state. 

\begin{figure}[t!]
\begin{center}
\includegraphics[width=0.35\textwidth, angle=270]{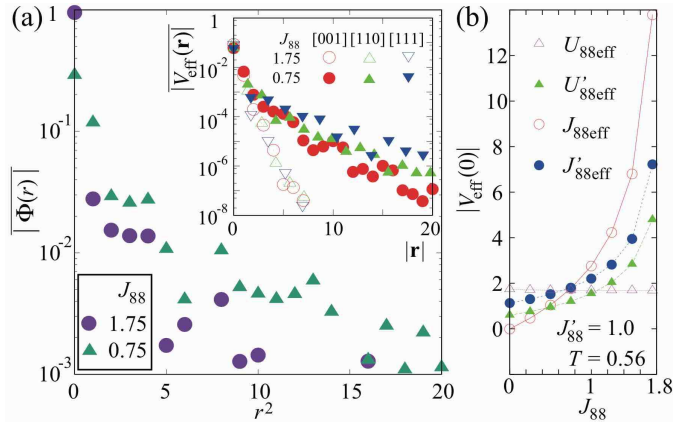}
\end{center}
\vspace{-0.5cm}
\caption{
(Color online) (a) Real space pair amplitude $\overline{|\Phi(r)|}$ as a function of the distance $r^2$ for ($J_{88}, J'_{88}$)=(0.75,1.0) and (1.75,1.0). 
Inset: $\overline{|V_{\rm eff}({\bf r})|}$ vs $|{\bf r}|$ for the direction parallel to the [001], [110], and [111] directions.
(b) $V_{\rm eff}(0)$ for $J'_{88}=1.0$ and $T=0.56$ as a function of $J_{88}$.  The effective interactions are evaluated in the normal state, ignoring the presence of $T_{\rm sc}$ 
(if it exists) for $T>0.56$.}
\label{fig-2}
\vspace{-1cm}
\end{figure}

To analyze the $d$-SC in more detail, the averaged real-space amplitude:
\begin{equation}
\overline{|\Phi(r)|}\equiv \frac{1}{Nn_{r}}\sum_{{\bf i}, {\bf j}}\sum_{\alpha\beta}|\langle \psi_{{\bf i}\alpha}\psi_{{\bf j}\beta}\rangle|\delta_{|{\bf i}-{\bf j}|,r},
\end{equation}
 for two sets of parameters of $J_{88}$ and $J'_{88}$ is shown in Fig. \ref{fig-2}(a), where $n_{r}$ is 
 the number of site-pairs with a distance $r$. For both $J_{88}=0.75$ and $1.75$, the order 
 parameter is largest for $r=0$,  
 and thus, it can be called the ``local'' Cooper pair. 
For larger $J_{88}$, the decay is 
 much faster. In the inset of Fig. \ref{fig-2}(a), the averaged effective interactions: 
 \begin{equation}
 \overline{|V_{\rm eff}({\bf r})|}\equiv \frac{1}{256}\sum_{\alpha\beta\gamma\delta} |V_{\rm eff}^{\alpha\beta\delta\gamma}({\bf r})|,\label{eqVr}
\end{equation}  
along the three symmetric directions [001], [110], and [111] are shown. In eq. (\ref{eqVr}), $V_{\rm eff}({\bf r})$  is the inverse-Fourier transform of eq. (\ref{Veff}). As expected from the behavior of $\overline{|\Phi(r)|}$, the effective interactions decay exponentially and the decay rate is higher for larger $J_{88}$.  
The profile of the local part of $V_{\rm eff}$ is key to gaining insight into the mechanism of $d$-SC. 
Figure \ref{fig-2}(b) demonstrates that 
 $J_{88{\rm eff}}$ and $J'_{88\rm eff}$ increase as $J_{88}$ increases. This stabilizes the f$^2$ $\G_3$ state, leading to the local Cooper pairs with $E_g$ irrep.

With regard to electron filling $n$ and the band parameter dependence of the phase diagram, it is shown that as $n$ decreases, all the ordered states including the SC are suppressed because of the decrease in the density of states on the FSs. For $U_{88}=2.0$ and $n=1.0$, there are no phases with broken symmetry in the calculations. $T_{\rm sc}$ for several sets of the band parameters are also examined. The results are qualitatively the same as those shown in Fig. \ref{fig-1}(a). This indicates that the band parameter details and any specific fluctuations play no role in realizing the $d$-SC discussed in this study.

%%%%%%%%%%%%%%%%%%%%%%%%%%% fig 2 %%%%%%%%%%%%%%%%%%%%%%%
%%%%%%%%%%%%%%%%%%%%%%%%%%%%%%%%%%%%%%%%%%%%%%%%%%%%%%%%%%

Our focus in this letter is on the 
$\G_3$ ground states in the f$^2$ configurations under cubic symmetry, which have been realized in e.g. Pr-based 1-2-20 compounds,
 and SC is found in various systems.\cite{PrU} Since the local nodal SC discussed in this letter does not require any specific fluctuation, if SC under high pressure far from the orbital ordered phase is realized,\cite{Matsubayashi} it is a good candidate for it. 
Another candidate is the classical heavy-fermion superconductor UBe$_{13}$.\cite{Ott} A possible ground state to explain the anomalous normal state is $\G_3$. It is suggested that there are point nodes around the [111] direction.\cite{Shimizu} 
The chiral $d$-wave state of local pairs described in this letter can provide
a key to understanding the enigmatic superconductivity of UBe$_{13}$. As a material 
with $\G_1$ ground states, it is argued that PrOs$_4$Sb$_{12}$ is an $s$-wave superconductor.\cite{PrOsSb}
This is indeed consistent with our theory and it is worth examining the heavy-fermion SC in our future studies.

In summary, we have demonstrated that local nodal superconductivity can appear in multiorbital systems. 
The nodal structures reflect the local two-electron ground state. 
We have also derived an effective model by local down folding and found  antiferro Hund's coupling and 
enhanced interorbital interactions, which are key to realizing local nodal superconductivity. 

%In summary, we have demonstrated a scheme of local down folding in multiorbital systems with SO interaction and 
%derived general low-energy effective Hamiltonian for $j=5/2$ multiplet in $O_h$ point group symmetry,
%which naturally explains the presence of the antiferro Hund's coupling and the enhanced interorbital interactions. Our scheme can be easily applied to other cases 
%with different multiplets and point group symmetry. We have also 
% analyzed  
%{\it local} and {\it nodal}
%superconductivity in multiorbital systems, focusing on cubic $\G_3$
%two-electron ground
%states in a simple toy model for $\G_8$ systems. Our main result is that 
%the two-electron CEF structure affects the renormalized {\it
%interactions} and leads to nodal superconductivity
%corresponding to the symmetry of the CEF ground state.  To proceed further, more
%elaborated numerical simulations are planned and investigating the full
%three-orbital model is necessary for discussing real materials.

{\it Acknowledgment}: 
This work was supported by a Grant-in-Aid for
Scientific Research [Grant Nos. 16H01079, 16H01081, 16H04017, and 16H04021] 
from the Japan Society for the Promotion of Science. T. N. was supported by RIKEN Special Postdoctoral Researchers Program.


\vspace{-0.7cm}




\begin{thebibliography}{99} %% The number "99" means that this list has more than nine items.
%\bibitem{Review} 
\bibitem{highTc} W. E. Pickett, Rev. Mod. Phys. {\bf 61}, 433 (1989).
\bibitem{SrRu} A. P. Mackenzie and Y. Maeno, Rev. Mod. Phys. {\bf 75}, 657 (2003). 
\bibitem{Fe} K. Ishida, Y. Nakai, and H. Hosono, J. Phys. Soc. Jpn. {\bf 78}, 062001 (2009).
\bibitem{heavy} G. R. Stewart, Rev. Mod. Phys. {\bf 73}, 797 (2001).
\bibitem{Miyake} K. Miyake, S. Schmidt-Rink, and C. M. Varma, Phys. Rev. B {\bf 34}, 6554(R) (1986), D. Scalapino, E. Loh, Jr., and J. E. Hirsch, Phys. Rev. B {\bf 34}, 8190(R) (1986).
\bibitem{He} A. J. Leggett, Rev. Mod. Phys. {\bf 47}, 331 (1975). 
\bibitem{Kittaka} S. Kittaka, Y. Aoki, Y. Shimura, T. Sakakibara, S. Seiro, C. Geibel, 
F. Steglich, H. Ikeda, and K. Machida, Phys. Rev. Lett. {\bf 112}, 067002 (2014).
%\bibitem{THattori} T. Hattori, et al., J. Phys. Soc. Jpn. {\bf 85}, 073711 (2016).
\bibitem{Nomoto} T. Nomoto, K. Hattori, and H. Ikeda, Phys. Rev. B {\bf 94}, 174513 (2016).
%Y. Tokura and N. Nagaosa, Science, \textbf{288}, 462	(2000).
\bibitem{Yotsuhashi} S. Yotsuhashi, H. Kusunose, and K. Miyake, J. Phys. Soc. Jpn. {\bf 71}, 389 (2002), K. Hattori, S. Yotsuhashi, and K. Miyake, J. Phys. Soc. Jpn. {\bf 74}, 839 (2005).
\bibitem{downfolding} F. Aryasetiawan, M. Imada, A Georges, G. Kotliar, S. Biermann, and A. I. Lichtenstein, Phys. Rev. B {\bf 70}, 195104 (2004). 
\bibitem{Hotta} T. Hotta and K. Ueda, Phys. Rev. B {\bf 67}, 104518 (2003).
\bibitem{PrU} See a recent review for Pr-based 1-2-20 compounds and references therein, T. Onimaru and H. Kusunose, J. Phys. Soc. Jpn. {\bf 85}, 082002 (2016).
\bibitem{Cox} D. L. Cox, Phys. Rev. Lett. {\bf 59}, 1240 (1987).
\bibitem{Hotta2} T. Hotta and H. Harima, J. Phys. Soc. Jpn. {\bf 75}, 124711 (2006).
\bibitem{core} C. J. Morningstar and M. Weinstein, Phys. Rev. D {\bf 54}, 4131 (1996). 
\bibitem{TraditionalNotation} $Wx=15B_{4}^0$ and $W(1-|x|)=180B_6^0$, where $B_{4,6}^0$ are the CEF parameter for $l=3$ orbitals.
See Ref. {\citenum{Hotta2}} and  K. R. Lea, M. J. M. Leask, and W. P. Wolf, J. Phys. Chem. Solids {\bf 23},
1381 (1962).
\bibitem{wavefunc} Definitions of the two-particle products:\\
$(\psi^\dag \psi^\dag)_{\G_1}$$=$$(a_\dw^\dag a_\up^\dag+b_\dw^\dag b_\up^\dag)/\sqrt{2}$, 
$(\psi^\dag \psi^\dag)_{\G_{5xy}}$$=$$(a_\dw^\dag b_\up^\dag+a_\up^\dag b_\dw^\dag)/\sqrt{2}$, \\
$(\psi^\dag \psi^\dag)_{\G_{5yz}}$$=$$(a_\dw^\dag b_\dw^\dag$$-$$a_\up^\dag b_\up^\dag)/\sqrt{2}$, 
$(\psi^\dag \psi^\dag)_{\G_{5zx}}$$=$$i(a_\dw^\dag b_\dw^\dag$$+$$a_\up^\dag b_\up^\dag)/\sqrt{2}$, \\
$(\psi^\dag \psi^\dag)_{\G_{3u}}$$=$$(a_\dw^\dag a_\up^\dag-b_\dw^\dag b_\up^\dag)/\sqrt{2}$, 
$(\psi^\dag \psi^\dag)_{\G_{3v}}$$=$$(a_\dw^\dag b_\up^\dag-a_\up^\dag b_\dw^\dag)/\sqrt{2}$, \\ 
$(\psi^\dag c^\dag)_{\G_{3u}}$$=$$(-a_\dw^\dag c_\up^\dag+a_\up^\dag c_\dw^\dag)/\sqrt{2}$, 
$(\psi^\dag c^\dag)_{\G_{3v}}$$=$$(b_\dw^\dag c_\up^\dag-b_\up^\dag c_\dw^\dag)/\sqrt{2}$, \\
$(\psi^\dag c^\dag)_{\G_{4x}}=[({\bf n}_{-}\psi_\dw^\dag) c_\dw^\dag-({\bf n}_{-}\psi_\up^\dag) c_\up^\dag]/\sqrt{2}$,\\
$(\psi^\dag c^\dag)_{\G_{4y}}=i[({\bf n}_{+}\psi_\dw^\dag) c_\dw^\dag+({\bf n}_{+}\psi_\up^\dag) c_\up^\dag]/\sqrt{2}$,\\
$(\psi^\dag c^\dag)_{\G_{4z}}$$=$$(a_\dw^\dag c_\up^\dag+a_\up^\dag c_\dw^\dag)/\sqrt{2}$, 
$(\psi^\dag c^\dag)_{\G_{5xy}}$$=$$(b_\dw^\dag c_\up^\dag+b_\up^\dag c_\dw^\dag)/\sqrt{2}$, \\
$(\psi^\dag c^\dag)_{\G_{5yz}}=[-({\bf m}_{-}\psi_\dw^\dag) c_\dw^\dag+({\bf m}_{-}\psi_\up^\dag) c_\up^\dag]/\sqrt{2}$,\\
$(\psi^\dag c^\dag)_{\G_{5zx}}=i[({\bf m}_{+}\psi_\dw^\dag) c_\dw^\dag+({\bf m}_{+}\psi_\up^\dag) c_\up^\dag]/\sqrt{2}$,\\
where ${\bf n}_{\pm}=(\cos\frac{2\pi}{3},\pm\sin\frac{2\pi}{3})$, 
${\bf m}_{\pm}=(-\sin\frac{2\pi}{3},\pm\cos\frac{2\pi}{3})$ and the product means $({\bf n}_{\pm}\psi^\dag_{\s})=
\cos\frac{2\pi}{3}a^\dag_\s\pm\sin\frac{2\pi}{3}b^\dag_\s$ and similarly $({\bf m}_{\pm}\psi^\dag_{\s})=
-\sin\frac{2\pi}{3}a^\dag_\s\pm\cos\frac{2\pi}{3}b^\dag_\s$. 
\bibitem{opRenormalization} In addition to the interaction part, one
	needs to renormalize {\it e.g.}, hopping terms.\cite{core} However, these 
	depend on the configurations and generate so-called correlated
	hopping terms. This cannot properly be
	taken into account in RPA calculations, while it is possible to
	compile them into {\it e.g.}, continuous-time quantum Monte Carlo for dynamical-mean-field analysis.
\bibitem{Hoshino2} Y. Nomura, S. Sakai, M. Capone, and R. Arita, Sci. Adv. {\bf 1}, 1500568 (2015), S. Hoshino and P. Werner, Phys. Rev. Lett. {\bf 118}, 177002 (2017).
 \bibitem{Koga} A. Koga and P. Werner, Phys. Rev. B {\bf 91}, 085108 (2015).
\bibitem{Hoshino} S. Hoshino and P. Werner, Phys. Rev. Lett. {\bf 115}, 247001 (2015).
\bibitem{dvec} $\epsilon({\bf k}) = -2t\sum_l c_l - 4t'\sum_l c_m c_n
 -8t''c_x c_y c_z$, $(l,m,n= $ cyclic),
$\vec{d}({\bf k}) =[d_z({\bf k}),d_x({\bf k})]=  \big[d(2c_z-c_x-c_y)+d'(2c_xc_y-c_yc_z-c_zc_x), \sqrt{3}d(c_x-c_y)+\sqrt{3}d'(c_yc_z-c_zc_x)\big],\ \ 
\eta_l({\bf k}) =(\eta+\eta' c_l) s_m s_n$, ($l,m,n=$ cyclic), where $t,\ t',\ t'',\ d,\ d',\ \eta,\ \eta'$ are parameters and $c_l=\cos k_l$ and $s_l=\sin k_l (l=x,y,z)$.
\bibitem{note} Note that the pair is local in terms of the local bases $a$ and $b$, while spreads in terms of the band bases.
\bibitem{Bishop} C. B. Bishop, G. Liu, E. Dagotto, and A. Moreo, Phys. Rev. B {\bf 93}, 224519 (2016).
\bibitem{Brydon} P. M. R. Brydon, L. Wang, M. Weinert, and D. F. Agterberg, Phys. Rev. Lett. {\bf 116}, 177001 (2016).
\bibitem{Takimoto} T. Takimoto, Phys. Rev. B {\bf 62},  R14641 (2000).
\bibitem{Takimoto2} T. Takimoto, T. Hotta, T. Maehira, and K. Ueda, J. Phys.: Condens. Matter {\bf 14}, L369-L375 (2002).
\bibitem{nonvanishV} $V^{\alpha\beta\delta\gamma}$ have nonvanishing elements only in $V_0^{1221}=V_0^{3443}=U_{88}$, $V_0^{1331}=V_0^{2442}=U^\prime_{88}+J_{88}/4$, $V_0^{1441}=V_0^{2332}=U^\prime_{88}-J_{88}/4$, 
$V_0^{1432}=V_0^{2341}=J_{88}/2$, and $V_0^{1243}=V_0^{3421}=J^\prime_{88}$. Note that $\G_0^{\alpha\beta\gamma\delta}=-\G_0^{\alpha\beta\delta\gamma}=
-\G_0^{\beta\alpha\gamma\delta}=\G_0^{\beta\alpha\delta\gamma}$.
\bibitem{NSR} P. Nozi\`{e}res and S. Schmitt-Rink, J. Low Temp. Phys. {\bf 59}, 195 (1985). 
\bibitem{d-BEC} G. Cao, L. He, and P. Zhuang, Phys. Rev. A {\bf 87}, 013613 (2013).
\bibitem{Matsubayashi} K. Matsubayashi, T. Tanaka, A. Sakai, S. Nakatsuji, Y. Kubo, and Y. Uwatoko, Phys. Rev. Lett. {\bf 109}, 187004 (2012).
\bibitem{Ott} H. R. Ott, H. Rudigier, Z. Fisk, and J. L. Smith, Phys. Rev. Lett. {\bf 50}, 1595 (1983).
\bibitem{Shimizu} Y. Shimizu, S. Kittaka, T. Sakakibara, Y. Haga, E. Yamamoto, H. Amitsuka, 
Y. Tsutsumi, and K. Machida, Phys. Rev. Lett. {\bf 114}, 147002 (2015).
\bibitem{PrOsSb} N. A. Frederick, T. D. Do, P.-C. Ho, N. P. Butch, V. S. Zapf, and M. B. Maple, Phys. Rev. B {\bf 69}, 024523 (2004).
\end{thebibliography}
\end{document}